\documentclass[final]{agujournal2019}
\usepackage{url} 
\usepackage{lineno}
\usepackage[inline]{trackchanges} 
\usepackage{soul}

\draftfalse

\journalname{JGR: Planets}

\begin{document}

%
%


\title{A Mid-to-Low Latitudes Field-Aligned Current (FAC) System at Jupiter Revealed by Juno MAG}

%
%




\authors{Wenyu Zhang\affil{1}, Omakshi Agiwal\affil{2}, Hao Cao\affil{1}, John E. P. Connerney\affil{3,4}, Krishan K. Khurana\affil{1}, Jeremy Bloxham\affil{5},  Ingo C. F. Müller-Wodarg\affil{6}, Scott J. Bolton\affil{7}}


\affiliation{1}{Department of Earth, Planetary and Space Science, University of California, Los Angeles (UCLA), Los Angeles, CA 90095, USA}
\affiliation{2}{Center for Space Physics, Boston University, MA 02215, USA}
\affiliation{3}{Space Research Corporation, Annapolis, MD 21403, USA}
\affiliation{4}{NASA Goddard Space Flight Center, Greenbelt, MD 20771, USA}
\affiliation{5}{Department of Earth and Planetary Sciences, Harvard University, Cambridge, MA 02138, USA}
\affiliation{6}{Department of Physics, Imperial College London, London SW7 2AZ, UK}
\affiliation{7}{Southwest Research Institute, San Antonio, TX, USA, 78238}




\correspondingauthor{Wenyu Zhang}{wenyuzhang@ucla.edu}
\correspondingauthor{Hao Cao}{hcao@epss.ucla.edu}



\begin{keypoints}

\item Juno MAG reveals large-scale residual $B_\varphi$ within 5~R$_\mathrm{J}$, with distinct spatial structures and a clear day--night asymmetry.

\item The residual $B_\varphi$ patterns suggest mid-to-low-latitude non-auroral interhemispheric field-aligned currents (IHFACs) at Jupiter.

\item Nightside IHFACs agree with JTIM thermosphere wind shear, whereas the dayside current system likely involves additional processes.

\end{keypoints}
%
%

%
%


\begin{abstract}
Large azimuthal magnetic field residuals observed by the Juno spacecraft within Jupiter’s innermost magnetosphere remain incompletely explained by existing magnetic field models. Using Juno MAG observations from the first 62 perijove passes, we investigate the residual magnetic field after removing the JRM33 internal field and the Con2020 magnetodisc field. We identify a pronounced day–night asymmetry in both the polarity and spatial organization of the residual azimuthal field. On the dayside, the residual field exhibits a coherent latitudinal ``sandwich’’ structure, characterized by opposite polarities at low-to-middle and higher latitudes, whereas the nightside distribution shows stronger longitudinal organization. Interestingly, the dayside polarity transition approximately follows the magnetic footprints of Jupiter’s main ring in certain longitude sectors. We then show that a local-time-dependent, mid-to-low-latitude IHFAC system with a current intensity of 0.47~MA~deg$^{-1}$ is broadly consistent with the observed large-scale $\Delta B_\varphi$ structure. Comparison with the Jupiter Thermosphere–Ionosphere Model (JTIM) shows that the nightside perturbations are broadly consistent with interhemispheric zonal wind shear, suggesting that thermospheric winds play an important role in generating the nightside IHFAC system. In contrast, the dayside residual field shows weaker agreement with the modeled wind shear, indicating that its origin may involve additional ionospheric electrodynamic processes. These findings reveal distinct local-time-dependent characteristics of Jupiter’s mid-to-low-latitude current system and provide new constraints on magnetosphere–ionosphere coupling in the innermost Jovian magnetosphere.
\end{abstract}

\section*{Plain Language Summary}
Jupiter possesses the strongest planetary magnetic field in the Solar System. Measurements from NASA's Juno spacecraft show that after accounting for Jupiter's known internal magnetic field and magnetosphere-ionosphere current systems, significant unexplained magnetic signals remain in measurements close to the planet. In this study, we analyze these residual magnetic signals and find that they exhibit a pronounced day--night asymmetry in both polarity and spatial organization. On the dayside, the magnetic field variations are primarily organized by latitude, with the boundary between regions of opposite magnetic polarity approximately following the magnetic footprints of Jupiter's main ring in certain regions. On the nightside, the magnetic field variations are more strongly organized by longitude. The magnetic field polarity also reverses between the dayside and nightside near the dawn terminator. We show that these magnetic field patterns are consistent with electric currents flowing along Jupiter's magnetic field lines between the northern and southern hemispheres at mid-to-low latitudes. Comparison with the latest thermosphere-ionosphere global circulation model suggests that the nightside current system is consistent with upper-atmospheric winds in the model, while the processes responsible for shaping the dayside structure remain uncertain and may involve additional solar-modulated electrodynamic processes.

%
%

%


%
%
%
%

\section{Introduction}

Jupiter hosts the strongest planetary magnetic field and the largest magnetosphere in the Solar System 
\cite{smith_planetary_1974,smith_jupiters_1975,ness_magnetic_1979,kivelson_magnetospheric_2004}. 
Internal plasma source and rotation-driven dynamics play an important role in this immense magnetosphere and establish the large-scale magnetosphere--ionosphere current systems 
\cite{hill_inertial_1979,dessler_plasma_1983,southwood_new_2001,khurana_configuration_2004}. 
The energy and momentum transfer between the ionosphere and magnetosphere of Jupiter is mediated by these field-aligned currents, ionospheric closure currents and radial current in the magnetodisk \cite<e.g.,>[]{cowley_origin_2001}.

Early in-situ magnetic field observations by Pioneer 10 \& 11 \cite{smith_planetary_1974,smith_jupiters_1975}, Voyager 1 \& 2 \cite{ness_magnetic_1979-1}, and later the Galileo orbiter \cite{kivelson_magnetospheric_2004,yu_galileo_2010} established that Jupiter’s magnetic field is dominated by a strong dipole tilted by approximately 9.6$^\circ$ with respect to the planet’s rotation axis, with an estimated dipole moment of order $1.5 \times 10^{20}$~T$\cdot$m$^3$. These earlier measurements provided critical constraints on Jupiter’s global magnetospheric dynamics \cite{dessler_plasma_1983,southwood_new_2001,kivelson_first_2003,kivelson_dynamical_2005,khurana_configuration_2004}. However, the limited spatial coverage at locations close to Jupiter of these early missions restricted detailed characterization of smaller-scale magnetic structures close to the planet.

The NASA Juno mission has fundamentally transformed our understanding of Jupiter’s magnetic field and plasma environment. Juno was placed into a highly eccentric polar orbit in 2016, enabling repeated close passes over the planet with periapsis distances as low as $\sim$1.008~R$_\mathrm{J}$ \cite{bolton_juno_2017}. The magnetometer (MAG) onboard provides high-precision vector magnetic field measurements, allowing both unprecedented resolution of Jupiter’s internal magnetic field and the identification of magnetic perturbations associated with magnetospheric-ionospheric current systems \cite{connerney_juno_2017}. These observations have revealed strong departures from a simple tilted dipole, including pronounced hemispheric asymmetry and localized magnetic anomalies \cite{connerney_new_2018,moore_complex_2018}. Notable examples include the "Great Blue Spot" (GBS), a compact region of intense radial magnetic flux located at low latitudes near System III East longitude of $\sim$270$^\circ$, and the Northern Hemisphere Flux Band (NHFB), a longitudinally extended band of enhanced outward radial magnetic field ($+B_r$) spanning a broad range of longitudes in the northern hemisphere \cite{connerney_new_2018,moore_complex_2018,connerney_new_2022,bloxham_differential_2022,bloxham_rapidly_2024}.

The latest Juno-era internal magnetic field model JRM33 \cite{connerney_new_2022} represents
Jupiter’s dynamo-generated magnetic field up to spherical harmonic degree and order 30, although individual harmonic coefficients are not fully resolved beyond degree 13. The model was derived from 30-sec averaged magnetic field measurements obtained during 32 orbits of the Juno Prime Mission. Comparing the field models derived from Juno measurements with the earlier Pioneer and Voyager magnetic field models, \citeA{moore_time_2019} demonstrated time-variations in Jupiter's dynamo-generated magnetic field on decadal time scales. Comparisons between JRM33 and its predecessor, the JRM09 model \cite{connerney_new_2018} further demonstrate measurable changes in Jupiter’s magnetic field over the first five years of Juno observations, providing further evidence of magnetic secular variation. These changes are particularly pronounced in the region of the Great Blue Spot, where ``rapid" temporal evolution of the magnetic field has been inferred from differences between JRM09 and JRM33 \cite{connerney_new_2022}. To more accurately account for the Jovian secular variation, \citeA{bloxham_differential_2022} developed the Differential Zonal Flux Rotation (DFR) model. In contrast to the static JRM33 model, the DFR model incorporates time dependence by jointly inverting the time-varying magnetic field and zonal flux velocity during the Juno observation period \cite{bloxham_differential_2022}. This approach revealed that the observed time evolution of the Great Blue Spot and other localized magnetic features is quantitatively consistent with differential rotation within Jupiter’s metallic hydrogen region. As a result, the DFR model captures the dynamical nature of Jupiter’s internal magnetic field and directly links the observed secular variation to interior flow patterns.

Magnetic enforcement of plasma corotation and centrifugal confinement at Jupiter lead to a large-scale magnetodisc structure extending from approximately 15 to beyond 50~R$_\mathrm{J}$, with a current sheet thickness of order 2--5~R$_\mathrm{J}$ \cite{connerney_modeling_1981,khurana_configuration_2004}. The associated azimuthal magnetic field is produced by the large-scale radial current system that transfers angular momentum from Jupiter to outward-moving plasma, thereby maintaining partial corotation throughout much of the middle magnetosphere. This current system produces magnetic perturbations of several hundred nanotesla (nT) near the magnetic equator and dominates the magnetic field morphology in the middle magnetosphere \cite{hill_inertial_1979,connerney_modeling_1981,connerney_comment_1981,khurana_configuration_2004,connerney_jovian_2020}. However, after accounting for the known internal and external magnetic field components, substantial residuals remains in the Juno MAG data, particularly within the inner magnetosphere ($<$5~R$_\mathrm{J}$). These residuals indicate that additional physical processes likely contribute to the observed magnetic field, in particular closer to the planet.

Field-aligned currents (FACs) and the associated closure currents in the magnetosphere and ionosphere are a fundamental component of magnetosphere--ionosphere coupling, transferring momentum and energy along magnetic field lines between the magnetosphere and the ionosphere. At Earth, extensive observational and theoretical studies have demonstrated the existence of mid- and low-latitude interhemispheric field-aligned currents driven by ionospheric wind dynamos, which close through Pedersen currents in both hemispheres \cite{fukushima_topics_1994,luhr_interhemispheric_2015,luhr_average_2019,buchert_entangled_2020}. More recently, magnetic field measurements from the Cassini Grand Finale revealed a comparable system of low-latitude interhemispheric currents at Saturn, which are thought to be generated by zonal wind shears across conjugate magnetic footprints in the thermosphere which produce azimuthal magnetic field $B_\varphi$ perturbations \cite{dougherty_saturns_2018,khurana_discovery_2018,hunt_currents_2019,provan_variability_2019,cao_landscape_2020,agiwal_constraining_2021}. These findings suggest that mid- and low-latitude interhemispheric field-aligned current systems may be a common feature of magnetized planets with dynamic ionospheres.

At Jupiter, FACs have been studied in the auroral region at high latitudes, where strong currents connect the magnetosphere to the ionosphere and contribute to intense auroral emissions \cite{cowley_origin_2001,grodent_jupiters_2003,nichols_magnetosphere-ionosphere_2004,ray_magnetosphere-ionosphere_2010}. In contrast, the existence and structure of mid- to low-latitude field-aligned currents at Jupiter remain poorly constrained. Given Jupiter’s rapid rotation, strong zonal winds \cite{muller-wodarg_temperatures_2025}, and highly conducting ionosphere \cite{nichols_magnetosphere-ionosphere_2003,agiwal_unraveling_2025,kurth_electron_2025}, analogous current systems may produce observable magnetic signatures in the azimuthal component of the magnetic field $B_\varphi$.

In this study, we present the spatial and temporal trends in the residual $B_\varphi$ observed by Juno in the inner magnetosphere, and investigate a wind-driven atmospheric dynamo as the possible driver. This study provides the first direct detection of a wind-driven ionospheric dynamo at Jupiter, as well as new insights into the nature of electrodynamics in Jupiter's upper atmosphere in the non-auroral region.

\section{Juno MAG Data}

\subsection{Juno Orbit and Observational Geometry}

\begin{figure*}[ht]
\centering
\includegraphics[width=0.6\textwidth]{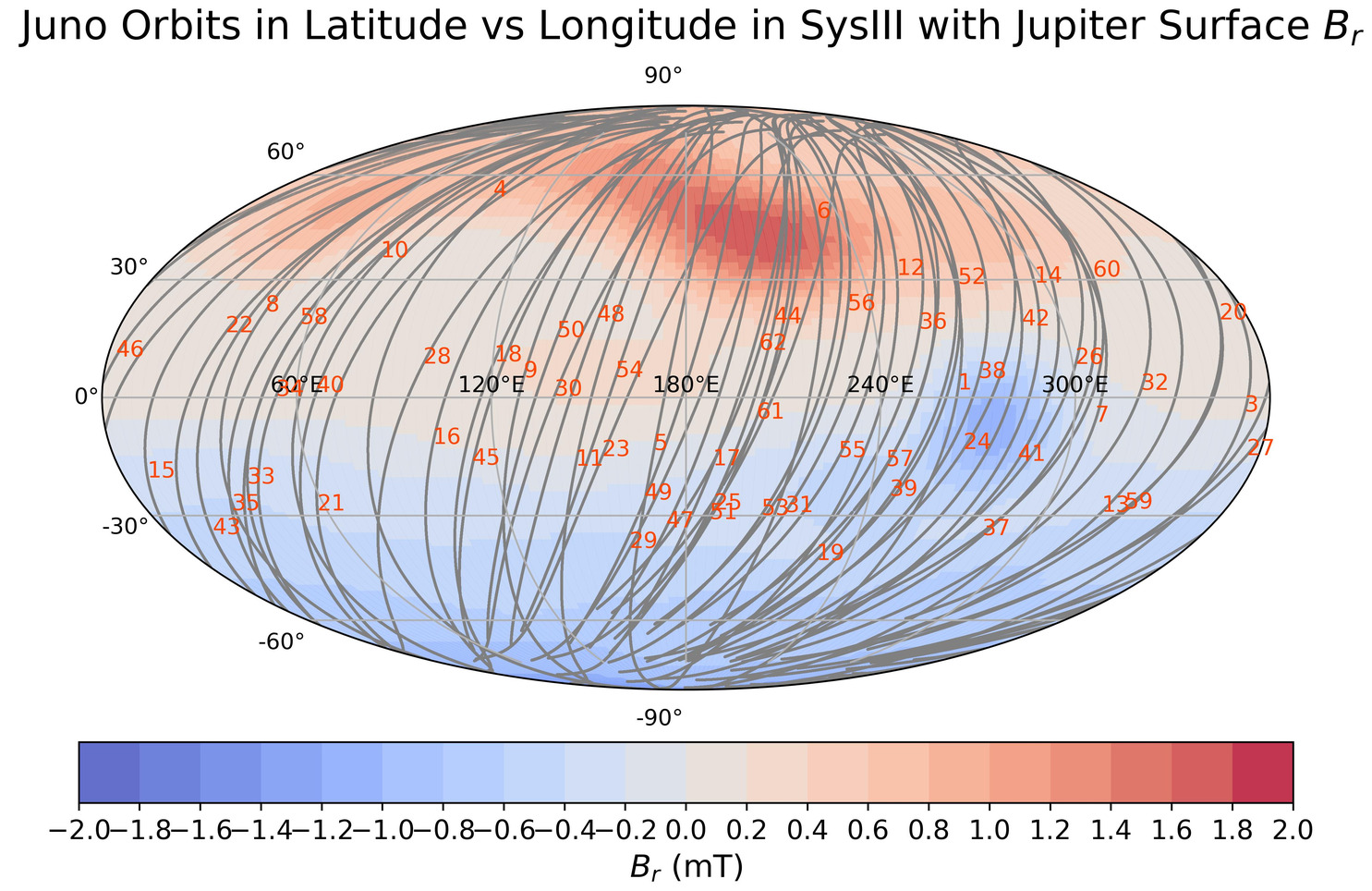}
\hfill
\includegraphics[width=0.55\textwidth]{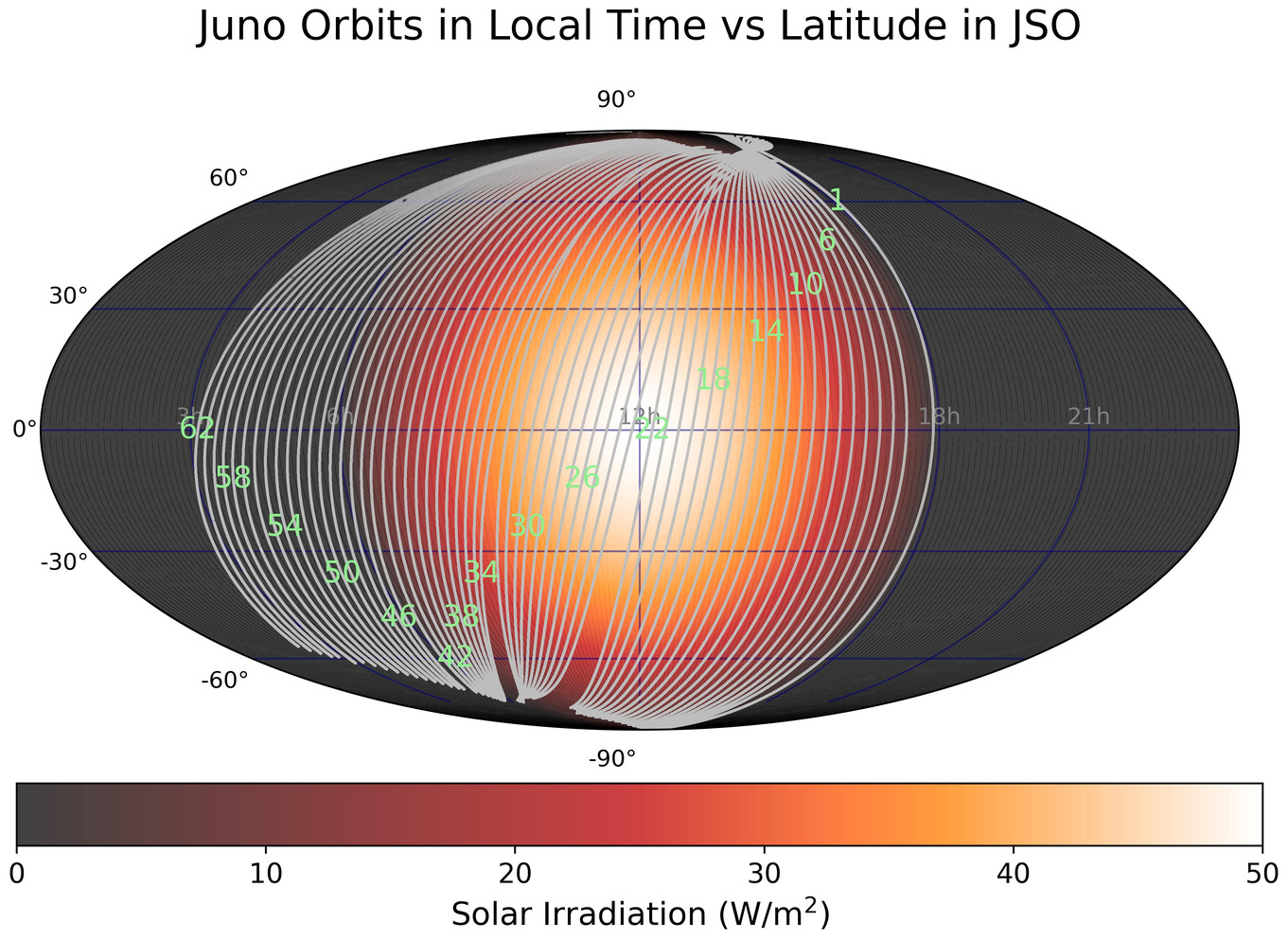}

\caption{
Spatial coverage of Juno orbits used in this study, shown in two complementary coordinate
systems. (a)~Orbital trajectories of the first 62 perijove passes (excluding PJ02) projected
in System~III longitude--latitude coordinates, overlaid on the radial magnetic field
component $B_r$ at the Jovian surface predicted by the JRM33 internal field model
\cite{connerney_new_2022}. (b)~The same set of orbits shown in the Jupiter--Sun--Orbit (JSO) frame, plotted in
local time versus latitude. The background shading represents the incident solar
irradiation at Jupiter, modeled as $F = F_0\,\mathrm{max}(\cos\phi \cos H, 0)$, where $\phi$
is latitude and $H$ is the local-time-dependent angle, highlighting the
day--night asymmetry in solar illumination. Together, the two panels illustrate both the
planetocentric (System~III) and local-time (JSO) sampling of Juno in the inner
magnetosphere, which forms the basis for the residual magnetic field analysis
presented in this study.
}
\label{fig:juno_orbit_coverage}
\end{figure*}

The NASA Juno spacecraft has been investigating Jupiter's internal structure, atmosphere,
magnetic field, magnetospheric dynamics, and moons and rings since 5 July 2016\cite{bolton_juno_2017}.
In this study, we analyze magnetic field observations from the first 62 perijove passes of the Juno mission (excluding PJ02). Each $\sim$53-day orbit carries Juno from a distant apojove at $\sim$113~R$_\mathrm{J}$ to perijove distance near $\sim$1.01~R$_\mathrm{J}$, corresponding to a closest altitude within a few thousand kilometers above the 1-bar pressure level. During each perijove, the spacecraft rapidly traverses from the northern to the southern hemisphere within 6.4 hours, sampling a broad range of longitudes and latitudes \cite{connerney_new_2018,connerney_new_2022}.

Figure~\ref{fig:juno_orbit_coverage} presents the spatial coverage of the Juno observations adopted in this work. Figure~\ref{fig:juno_orbit_coverage}a shows the orbital trajectories projected in System~III longitude--latitude coordinates, overlaid on the map of the radial magnetic field component, $B_r$, at the Jovian surface predicted by the JRM33 internal field model \cite{connerney_new_2022}. System III coordinate rotates with Jupiter and is therefore suitable for studying the field originating inside Jupiter. The polar orbit geometry and repeated perijove sampling enable Juno to cover a broad range of longitudes and latitudes, including regions associated with strong non-dipolar magnetic anomalies such as the GBS and NHFB. The same set of orbits is shown in Figure~\ref{fig:juno_orbit_coverage}b in the Jupiter--Sun--Orbit (JSO) coordinate system, where the X-axis points from Jupiter toward the Sun, the Y-axis lies in Jupiter's orbital plane and points opposite to its orbital motion, and the Z-axis completes the right-handed coordinate system \cite{wang_jupiters_2023}. This coordinate system is suitable for evaluating local time dependent processes in the Jovian system. Due to the slow precession of Juno’s orbital plane over the course of the mission, the local time of perijove gradually moves from the dusk sector ($\sim$18~h LT) in earlier orbits toward the dawn sector ($\sim$6~h LT) and into the pre-midnight sector in later orbits. As a result, the dataset provides extensive local-time coverage at low and mid latitudes, which is essential for identifying systematic day--night asymmetries in the residual magnetic field.

\subsection{Amplitude of the Residual Magnetic Field}

\begin{figure}[ht]
\centering
\includegraphics[width=0.9\linewidth]{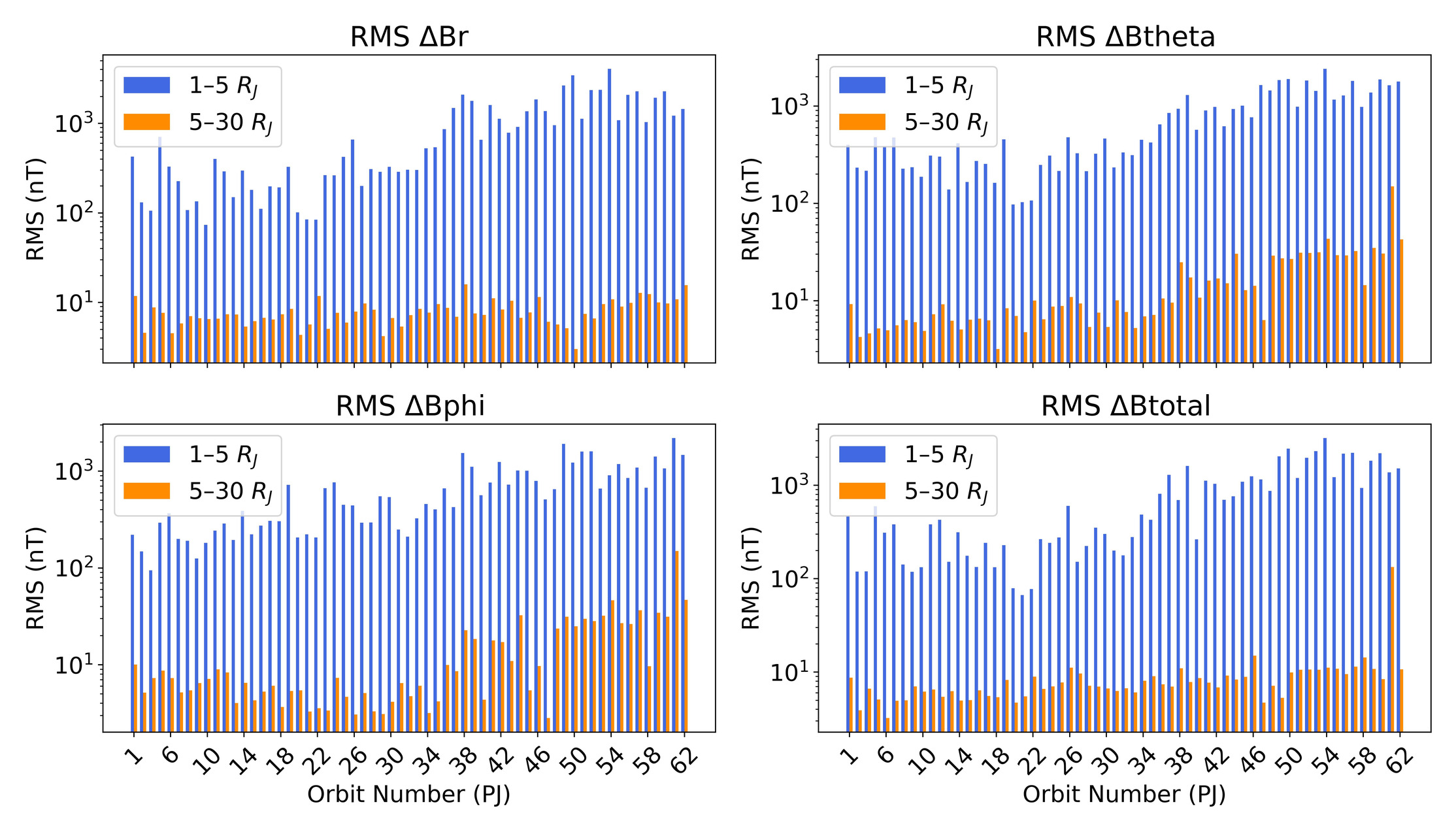}
\caption{Root-mean-square (RMS) of residual magnetic field components and total field per orbit, computed from Juno MAG data after removing the JRM33 \cite{connerney_new_2022} and Con2020 \cite{connerney_jovian_2020} models. Blue and orange bars denote the inner (1--5~R$_\mathrm{J}$) and middle (5--30~R$_\mathrm{J}$) magnetospheric regions.}
\label{fig:RMS_compare}
\end{figure}

To quantify the deviation of Juno's measured magnetic field from existing models, we computed the root--mean--square (RMS) amplitude of the residual field for each orbit. The residual magnetic field is defined as
\begin{equation}
\Delta \mathbf{B} = \mathbf{B}_\mathrm{data} - \mathbf{B}_\mathrm{JRM33} - \mathbf{B}_\mathrm{Con2020},
\label{eq:Bres}
\end{equation}
where $\mathbf{B}_\mathrm{data}$ is the magnetic field measured by Juno MAG \cite{jep_connerney_juno_2024}, $\mathbf{B}_\mathrm{JRM33}$ is the internal field based on the JRM33 model \cite{connerney_new_2022}, and $\mathbf{B}_\mathrm{Con2020}$ is the external magnetodisc field from the Con2020 magnetodisk model \cite{connerney_jovian_2020} with default parameters from \citeA{wilson_internal_2023}. For each orbit, the RMS value of the residual magnetic field components ($B_r$, $B_\theta$, $B_\varphi$) and total field magnitude was calculated as
\begin{equation}
\mathrm{RMS} \, \Delta B = 
\sqrt{\frac{1}{N} \sum_{i=1}^{N} 
\left( B_{i,\mathrm{data}} - B_{i,\mathrm{JRM33}} - B_{i,\mathrm{Con2020}} \right)^2 },
\label{eq:RMS}
\end{equation}
where $N$ is the number of data points within the selected radial range for a given orbit. The RMS residuals were evaluated separately for two radial ranges: 1--5~R$_\mathrm{J}$ (inner magnetosphere) and 5--30~R$_\mathrm{J}$ (middle magnetosphere). The results are shown in Figure~\ref{fig:RMS_compare}. In all components, the residual field within 1--5~R$_\mathrm{J}$ (blue bars) is one to two orders of magnitude larger than that between 5--30~R$_\mathrm{J}$ (orange bars), indicating substantial unmodeled magnetic field close to Jupiter. The RMS residuals exhibit clear orbit-to-orbit variability, with values up to several thousand nT along certain perijove passes.

\subsection{Spatial Distribution of the Residual Magnetic Field}

\begin{figure*}[ht]
\centering
\includegraphics[width=\textwidth]{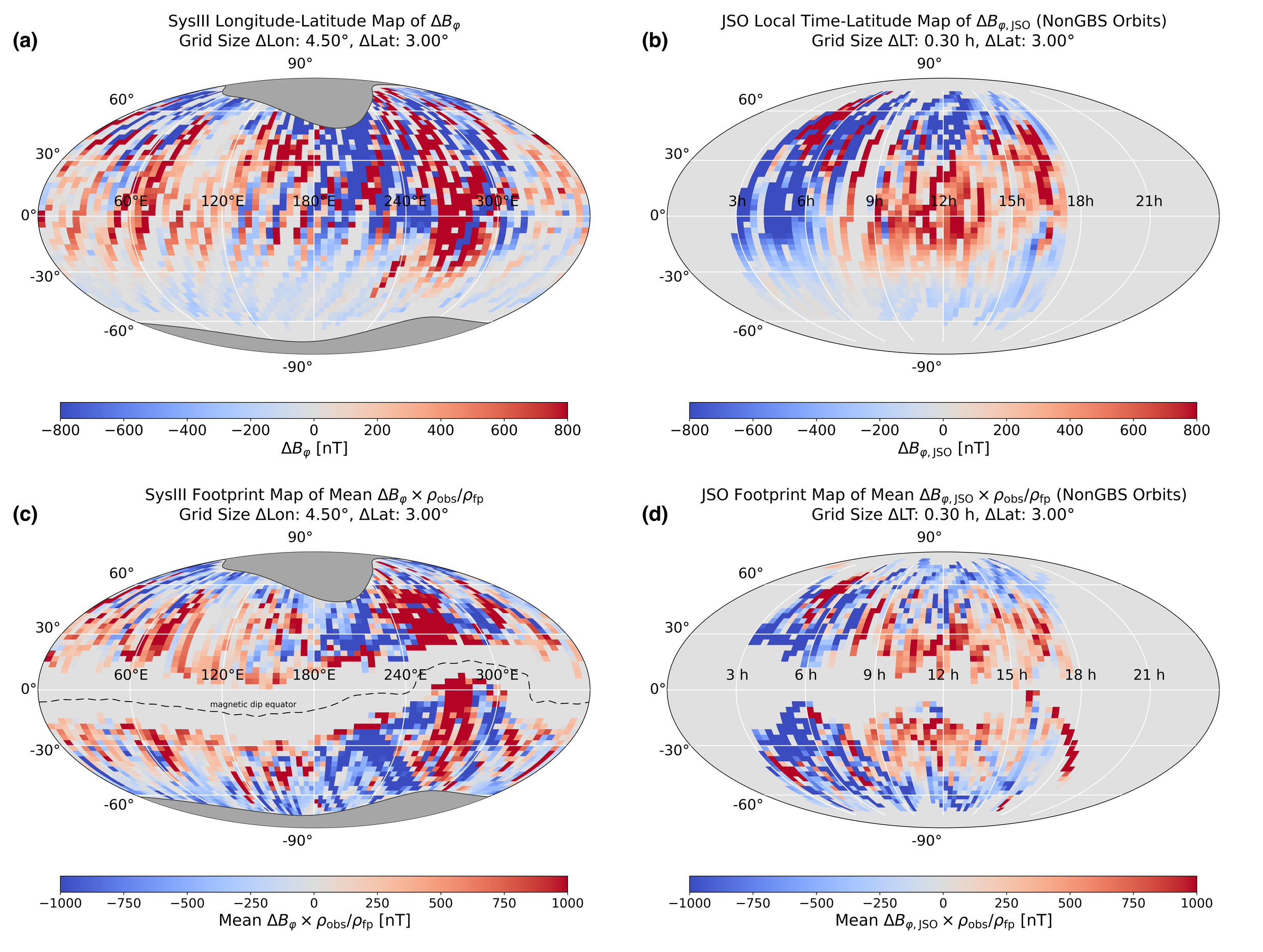}
\caption{Global spatial distributions of the azimuthal residual magnetic field $\Delta B_\varphi$ within 1–5R$\mathrm{J}$. (a) In-situ $\Delta B_\varphi$ in System III longitude–latitude coordinates. (b) In-situ $\Delta B_{\varphi,\mathrm{JSO}}$ in JSO local time–latitude coordinates for non-GBS orbits. (c) System~III ionospheric footprint map of the normalized mean $\Delta B_\varphi$. (d) Corresponding JSO ionospheric footprint map for non-GBS orbits. The footprint maps are normalized by $\rho_{\mathrm{obs}}/\rho_{\mathrm{fp}}$. Gray regions indicate areas poleward of the Io footprint, and the dashed black curve marks the magnetic dip equator.}
\label{fig:Bphi_mollweide}
\end{figure*}

\begin{figure}[ht]
\centering
\includegraphics[width=\linewidth]{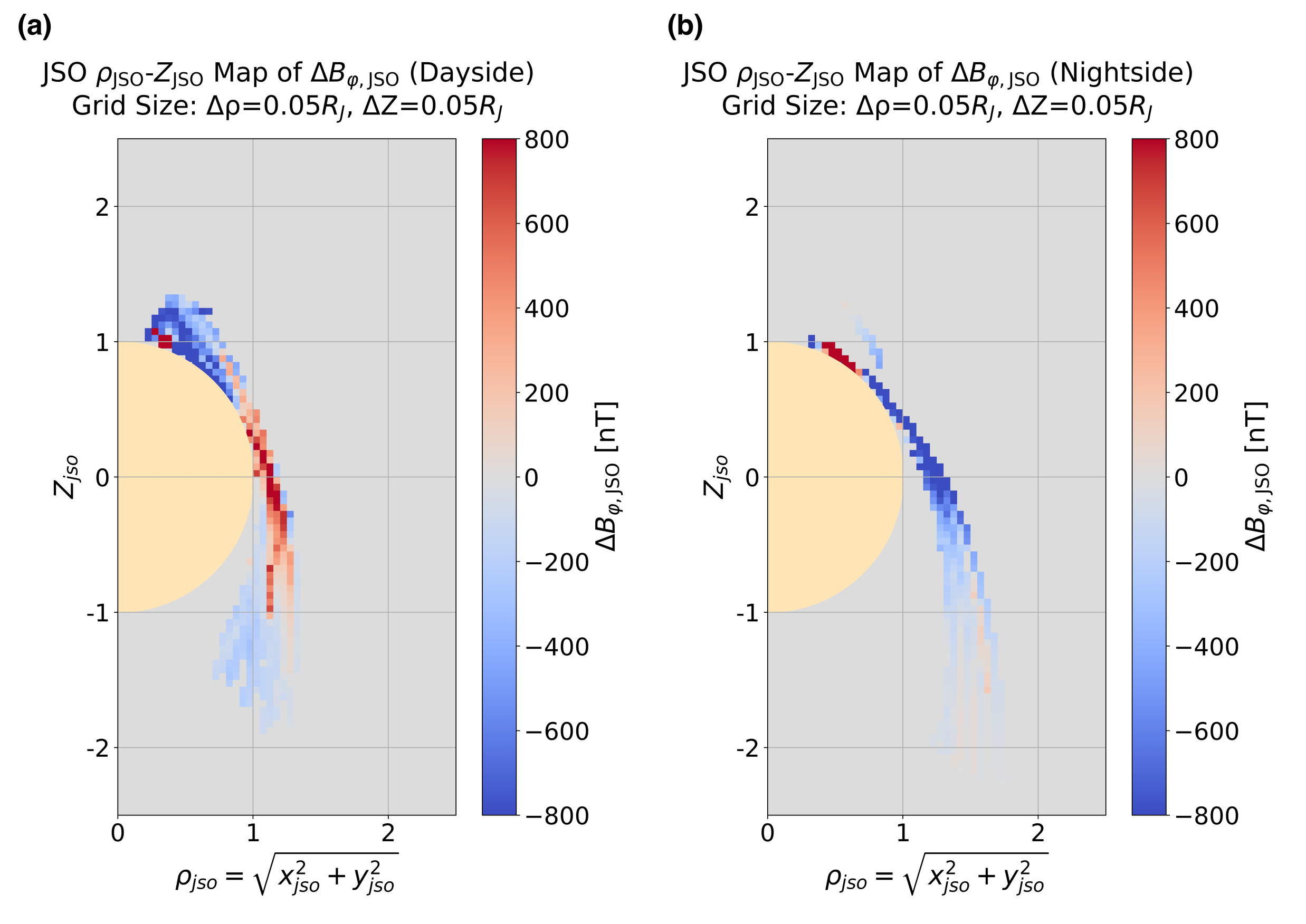}
\caption{Cylindrical $\rho_{\mathrm{JSO}}$–$Z_{\mathrm{JSO}}$ distributions of the azimuthal residual magnetic field $\Delta B_{\varphi,\mathrm{JSO}}$ on the (a) dayside and (b) nightside. The two sectors exhibit opposite low- to mid-latitude magnetic field polarities.}
\label{fig:Bphi_zrho}
\end{figure}

\begin{figure*}[ht]
\centering
\includegraphics[width=\textwidth]{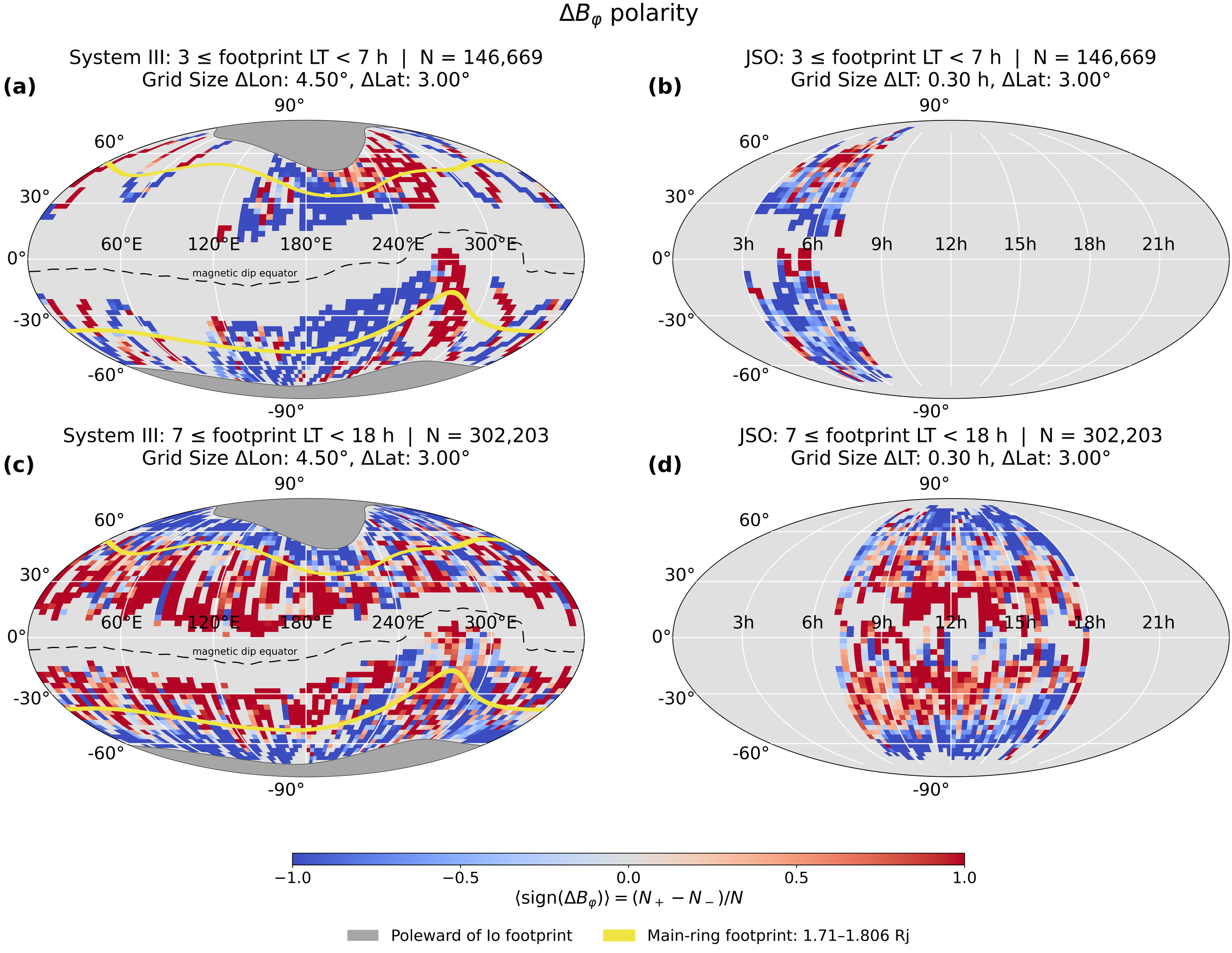}
\caption{Local-time dependence of the ionospheric footprint polarity distribution of $\Delta B_\varphi$. (a,b) Nightside sector (3–7h LT) in (a) System III longitude–latitude and (b) JSO local time–latitude coordinates. (c,d) Dayside sector (7–18h LT) in (c) System III and (d) JSO coordinates. Colors represent the normalized net sign count, $(N_+-N_-)/N$, where $N_+$ and $N_-$ are the numbers of positive and negative $\Delta B_\varphi$ measurements in each bin, respectively, and $N=N_++N_-$. Gray regions indicate areas poleward of the Io footprint, while yellow curves mark the magnetic footprints of Jupiter’s main ring (1.71–1.806~R$_\mathrm{J}$).}
\label{fig:Bphi_LT_sectors}
\end{figure*}

We next examine the spatial organization of the azimuthal residual magnetic field $\Delta B_\varphi$ observed within 1–5R$_\mathrm{J}$. We restrict our analysis to data that maps to latitudes below Io's footprint in the ionosphere, as Jupiter's magnetic field is strongly altered by the Io plasma torus and other magnetospheric processes outside this region, including auroral FACs.

Figure \ref{fig:Bphi_mollweide} presents the residual field distributions in both the System III and JSO coordinate systems, including the in-situ measurements and their corresponding ionospheric footprint projections. Overall, the residual field exhibits pronounced longitudinal variations in System III coordinates, while the JSO distributions reveal a clear day--night polarity reversal and a latitudinal ``sandwich'' structure on the dayside, with opposite polarities at low-to-middle and higher latitudes. These features are also evident in the corresponding ionospheric footprint maps, indicating that the large-scale magnetic perturbations exhibit coherent spatial organization across both hemispheres.

In the System III frame (Figure~\ref{fig:Bphi_mollweide}a), strong localized residuals exceeding 800~nT are observed between 240$^\circ$ and 300$^\circ$E, corresponding to perijove passes near the Great Blue Spot (GBS). Additional enhanced residuals occur at northern middle latitudes between 180$^\circ$ and 240$^\circ$E, near the northern hemisphere flux band (NHFB). These enhancements may be associated with the fine spatial structure and secular evolution of the GBS \cite{bloxham_differential_2022}, or due to ionospheric dynamics in a strong magnetic field region. The generally larger residual amplitudes in the northern hemisphere are partly attributable to Juno’s lower spacecraft altitude during northern hemisphere passes in later orbits.

To examine the spatial distribution of these perturbations in the ionosphere, we map each in-situ $\Delta B_\varphi$ measurement onto its northern and southern ionospheric footprints using the JRM33 magnetic field model (Figure~\ref{fig:Bphi_mollweide}c). The mean residual field in each spatial bin is normalized by the cylindrical radial distance ratio $\rho_{\mathrm{obs}}/\rho_{\mathrm{fp}}$ to account for the difference between the spacecraft observation location and its ionospheric footprint. The resulting System~III footprint map reveals substantial longitudinal variations in the residual field polarity. Predominantly positive $\Delta B_\varphi$ at low and middle latitudes, accompanied by negative values at higher latitudes, is observed between 0$^\circ$ and 180$^\circ$E. In contrast, negative residuals dominate much of the 180$^\circ$--240$^\circ$E sector, while positive values prevail over a broad range of latitudes between 240$^\circ$ and 300$^\circ$E.

However, Juno's orbital coverage does not uniformly sample all longitudes and latitudes at every local time. Consequently, longitudinal variations in the globally averaged residual field may be influenced by local-time-dependent structures. To address this ambiguity, we further examine the residual field in the JSO frame and separate the ionospheric footprint measurements by local-time sector, allowing us to compare the spatial organization of $\Delta B_\varphi$ under different local-time conditions.

To assess the local-time dependence of the observed residual field patterns, we first examine the $\Delta B_\varphi$ distributions in the JSO frame, excluding orbits near the GBS longitude sector (Figure~\ref{fig:Bphi_mollweide}b and \ref{fig:Bphi_mollweide}d; see Appendix~A for the orbit selections). The JSO distributions reveal a pronounced local-time dependence, with predominantly positive $\Delta B_\varphi$ at low and middle latitudes on the dayside and negative values near dawn. At higher latitudes, the polarity generally reverses, producing the characteristic ``sandwich’’ structure on the dayside. This day–night polarity reversal is also evident in the cylindrical $\rho_{\mathrm{JSO}}$–$Z_{\mathrm{JSO}}$ projection (Figure~\ref{fig:Bphi_zrho}). On the dayside (Figure~\ref{fig:Bphi_zrho}a), positive residuals dominate near the magnetic equatorial region, whereas negative residuals are observed at higher northern latitudes. On the nightside (Figure~\ref{fig:Bphi_zrho}b), negative residuals dominate at low and middle latitudes, accompanied by localized positive perturbations at higher northern latitudes. Although the observed amplitudes are influenced by differences in Juno’s orbital geometry and radial sampling between the two sectors, the opposite low- to mid-latitude polarities remain clearly distinguishable.

Beyond this polarity reversal, the residual field exhibits distinct spatial organization between the dawn and dayside sectors. To investigate this difference, we separate the ionospheric footprint measurements into 3–7h LT and 7–18h LT intervals and calculate the normalized net sign count of $\Delta B_\varphi$ in each spatial bin (Figure~\ref{fig:Bphi_LT_sectors}). Unlike the mean residual field maps, this quantity emphasizes the dominant polarity independently of the residual amplitude. In the dawn sector (Figure~\ref{fig:Bphi_LT_sectors}a and \ref{fig:Bphi_LT_sectors}b), the polarity distribution exhibits pronounced longitudinal organization in System III coordinates, with alternating positive and negative regions across different longitude sectors. In contrast, the dayside distribution (Figure~\ref{fig:Bphi_LT_sectors}c and \ref{fig:Bphi_LT_sectors}d) is predominantly organized by latitude, exhibiting positive polarity at low and middle latitudes and negative polarity at higher latitudes in both hemispheres. Interestingly, the dayside polarity transition approximately follows the mapped magnetic footprint of Jupiter’s main ring (1.71–1.806~R$_\mathrm{J}$) over certain longitude sectors, particularly in the southern hemisphere. While this spatial correspondence does not establish a causal relationship, it raises the possibility of a connection between the observed residual magnetic field structure and ring–planet electrodynamic coupling.

We further examine the $\Delta B_\varphi$ polarity at comparable magnetic footprints and local times across different perijove passes. Juno only traverses field lines with $M_{\mathrm{shell}}<1.8~R_\mathrm{J}$ only at radial distances $r<1.5~R_\mathrm{J}$; therefore, this comparison is restricted to mid-latitude observations ($1.8~R_\mathrm{J}<M_{\mathrm{shell}}<5~R_\mathrm{J}$). On the dayside, observations closer to Jupiter ($r<1.5~R_\mathrm{J}$) exhibit predominantly negative polarity across longitudes, whereas those at larger radial distances ($r>1.5~R_\mathrm{J}$) retain a more longitudinally structured distribution. In contrast, the nightside polarity remains relatively consistent across radial distances. Due to Juno's gradual orbital precession in local time, observations mapping to similar ionospheric footprints at comparable local times are often separated by substantial time intervals (months to years). Thus, the observed dayside differences may reflect radial variations or secular evolution of the mid-latitude current system.

Taken together, these observations reveal a pronounced day–night asymmetry in both the polarity and spatial organization of the azimuthal residual magnetic field. The predominantly longitudinal organization near dawn contrasts with the latitudinal ``sandwich’’ structure on the dayside, suggesting that the relative contributions of different electrodynamic processes may vary with local time. The possible correspondence between the dayside polarity transition and the main-ring magnetic footprint provides an additional observational constraint for understanding the origin of these perturbations.

\section{Estimating the amplitude of the inter-hemispherical field-aligned current system}
\begin{figure}[t]
\centering
\includegraphics[width=\textwidth]{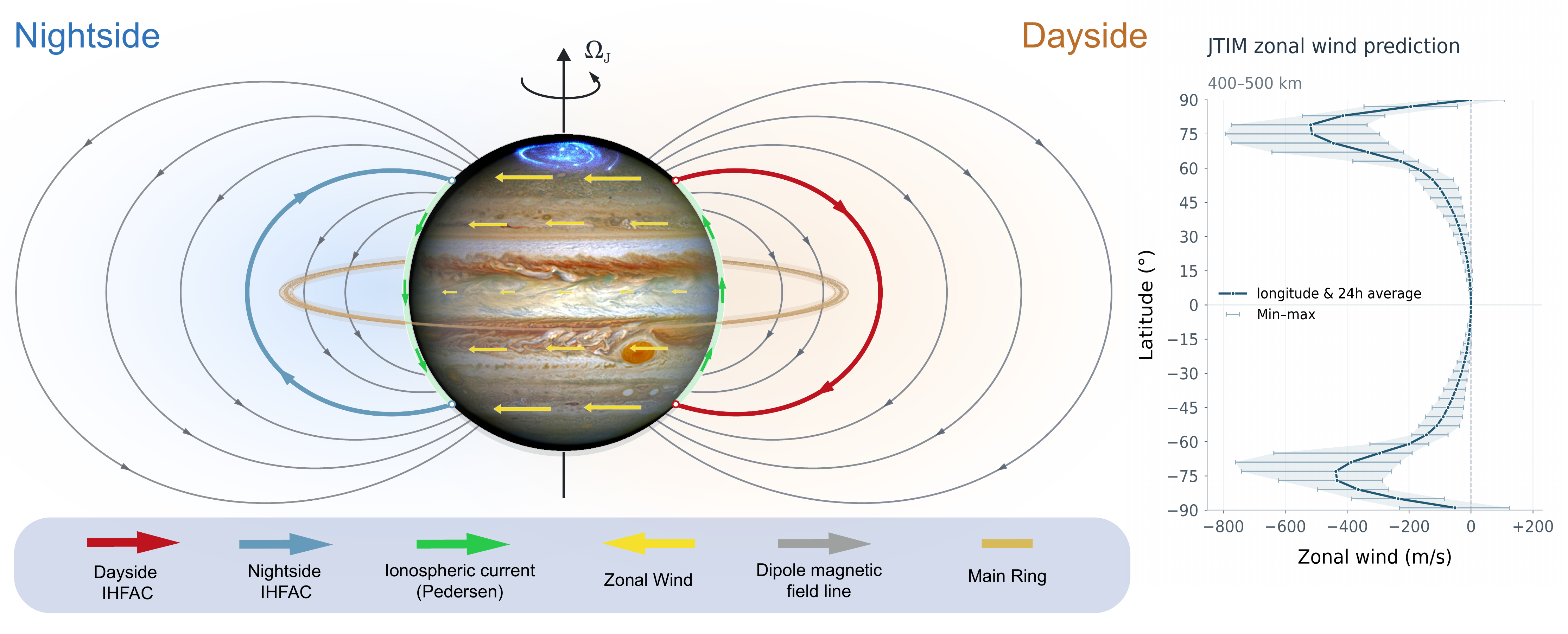}
\caption{Schematic illustration of the proposed interhemispheric field-aligned current (IHFAC) system and its relationship to thermospheric zonal winds. Red and blue loops denote the dayside and nightside IHFAC systems, respectively, with their opposite current directions, while green segments represent the associated Pedersen current closure in the Jovian ionosphere. Yellow arrows illustrate the zonal winds, and gray curves represent dipole magnetic field lines. The main ring is shown in brown. The right panel shows the latitudinal profile of the JTIM-predicted zonal wind at 400--500~km altitude, illustrating the wind structure used for comparison with the inferred IHFAC system. The background image of Jupiter is adapted from the NASA Hubble mission team.
}
\label{fig:IHFAC_model_sketch}
\end{figure}

In this section we describe the observed azimuthal residual magnetic field as being produced by a system of interhemispheric field-aligned currents (IHFACs). A schematic illustration of the proposed current geometry is shown in Figure~\ref{fig:IHFAC_model_sketch}. Unlike the large-scale corotation-enforcement current system, whose azimuthal magnetic perturbations arise from radial current flow in the middle magnetosphere and FACs and ionospheric closure current\cite<e.g.,>[]{connerney_new_2022}, the current system considered here is confined to the inner magnetosphere and consists of interhemispheric field-aligned currents closing through Pedersen currents in the Jovian ionosphere.The current geometry is parameterized by a single spatial parameter, $L_{JSO}$ (in units of $R_J$), which specifies the equatorial crossing distance of the current-carrying flux tubes. The current system is defined in the JSO coordinate and separated into dayside and nightside sectors based on local time. The dayside spans 6.5--17.5 LT and the nightside spans 18.5--5.5 LT (cross midnight). The current system is discretized into 80 field-aligned current loops, uniformly distributed in local time on the JSO equatorial plane at $r= L_{JSO}$, with equal numbers assigned to the dayside and nightside sectors. Magnetic field lines are traced in System III coordinates using the JRM33 internal field model \cite{connerney_new_2022,wilson_internal_2023}. For each orbit, the epoch near periapsis is used to determine the System III configuration, and seed locations defined in JSO are transformed into System III prior to field-line tracing. Along each field line (Figure~\ref{fig:IHFAC_model_sketch}), the current flows between the northern and southern hemispheres. On the dayside, the current is directed from north to south, and on the nightside the direction is reversed. Current closure is imposed through ionospheric paths on a spherical shell at $r=1$~R$_\mathrm{J}$. The resultant magnetic field is computed using the Biot--Savart law. The current is uniformly distributed among all loops. The resulting magnetic field is transformed back into the JSO frame for comparison with observations.

Given linearity of the Biot--Savart law, the modeled azimuthal $\Delta B_{\varphi,\mathrm{model}}$ scales linearly with the current amplitude. For each orbit, the optimal current amplitude is obtained by scaling once the loop geometry is specified. The fitting procedure is restricted to data from low- and mid-latitude non-GBS orbits ($|\lambda_{\mathrm{JSO}}| < 60^\circ$). The best-fit $L_{JSO}$ is determined using a shape-based metric defined by the normalized zero-lag correlation between modeled and observed $\Delta B_\varphi$.

\begin{figure*}[ht]
\centering
\includegraphics[width=\textwidth]{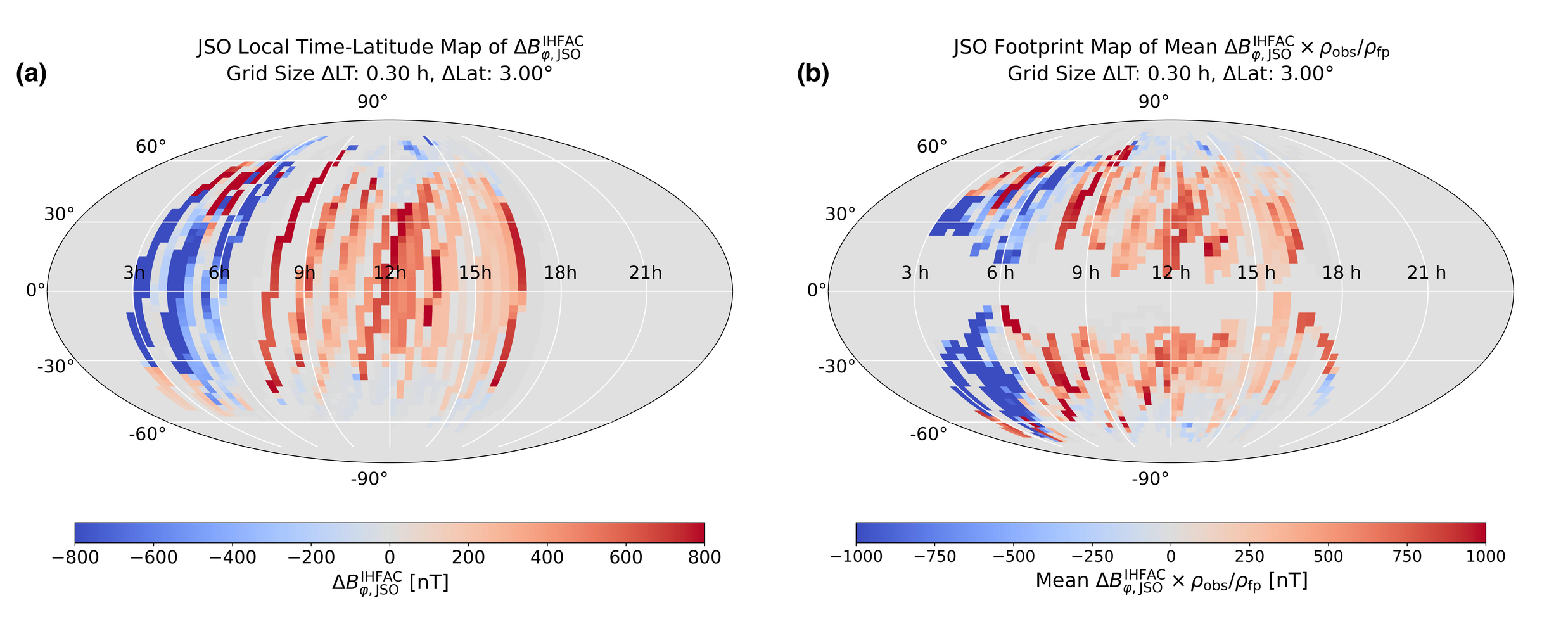}
\caption{(a) IHFAC modeled $\Delta B_{\varphi,\mathrm{JSO}}$ in local time--latitude coordinates, assembled from fitted perijove predictions. (b) Corresponding normalized mean modeled ionospheric footprint map, weighted by $\rho_{\rm obs}/\rho_{\rm fp}$. The fits use non-GBS data at $|\lambda_{\rm JSO}|<60^\circ$; see Appendix A for excluded perijoves.}
\label{fig:IHFAC_global}
\end{figure*}

The global prediction of the model is shown in Figure~\ref{fig:IHFAC_global}a, with the corresponding mapped ionospheric footprint distribution in Figure~\ref{fig:IHFAC_global}b. Despite being constrained only by low- and middle-latitude data ($|\lambda|<60^\circ$), the model reproduces the key large-scale features observed in the residual field, including the day--night asymmetry and the latitudinal ``sandwich'' structure. For the perijove passes with non-boundary fitted current amplitudes, the inferred current has a median value of $0.47~\mathrm{MA~degree^{-1}}$, corresponding to total currents of $154.9~\mathrm{MA}$ over the modeled local-time extent, which is comparable to the total current flowing in the magnetodisk current system.  The fitted current shells have a mean $L_{JSO}$ shell of $2.60$~R$_\mathrm{J}$ and a median value of $2.29$~R$_\mathrm{J}$. 

This exercise suggests that a low-latitude interhemispheric current system with amplitude comparable to that of the magnetodisk current is broadly consistent with the observed $\Delta B_\varphi$. Not all orbits are well reproduced. A subset of perijove passes (see Appendix~A) require more spatially complex current structures.

\section{Comparison with the latest Jupiter-Thermosphere-Ionosphere Model~(JTIM)}
At Earth, sub-auroral magnetic field perturbations are linked to a wind-driven atmospheric dynamo which drives inter-hemispheric field aligned currents as a part of the Sq-current system \cite{richmond_ionospheric_1979,fukushima_topics_1994,yamashita_seasonal_2002}, which are highly structured in local time. At Saturn, sub-auroral magnetic field perturbations \cite{dougherty_saturns_2018} have been linked to an analogous wind-driven thermospheric dynamo \cite{khurana_discovery_2018} as well as ring-planet coupling \cite{waite_chemical_2018,agiwal_current_2024}.
Jupiter has a substantial ionosphere \cite{stallard_identification_2018,melin_ionospheric_2024,kurth_electron_2025}, thin dusty rings \cite{showalter_jupiters_1987,ockert-bell_structure_1999}, and recent studies suggest strong thermospheric winds \cite{odonoghue_global_2021,muller-wodarg_temperatures_2025}. Similar processes to those observed at Earth and Saturn could also be altering Jupiter's magnetic field internal to the Io plasma torus.

Given the inter-hemispherical nature of the inferred current system, a thermosphere wind-driven scenario like that inferred at Saturn seems to be a reasonable one \cite{khurana_discovery_2018,hunt_currents_2019}. Recent thermosphere--ionosphere models (e.g., JTIM) predict predominantly westward zonal winds in Jupiter’s upper atmosphere, 
with typical velocities of $\sim$850 m·s$^{-1}$ outside the auroral regions.  These winds are driven by magnetospheric coupling and angular momentum transfer,  with ion drag—estimated to be an order of magnitude stronger than at Saturn—playing  a key role in regulating the global circulation and enabling pole-to-equator energy transport \cite{muller-wodarg_temperatures_2025}.

In this framework, the polarity of the observed $\Delta B_\varphi$ provides a constraint on the direction of the field-aligned current and therefore on the relative zonal wind shear, $\Delta w$, between conjugate hemispheres, with $\mathbf{E}_{Drive}=-\mathbf{u}\times\mathbf{B}$ as the initial (kinematic) driver. Such a connection has been established at Saturn, where interhemispheric currents with amplitude of $0.02~\mathrm{MA~degree^{-1}}$ are driven by differences in zonal winds between conjugate ionospheres \cite{dougherty_saturns_2018,khurana_discovery_2018,hunt_currents_2019,provan_variability_2019,cao_landscape_2020,agiwal_constraining_2021}. 

\begin{figure*}[ht]
\centering
\includegraphics[width=\textwidth]{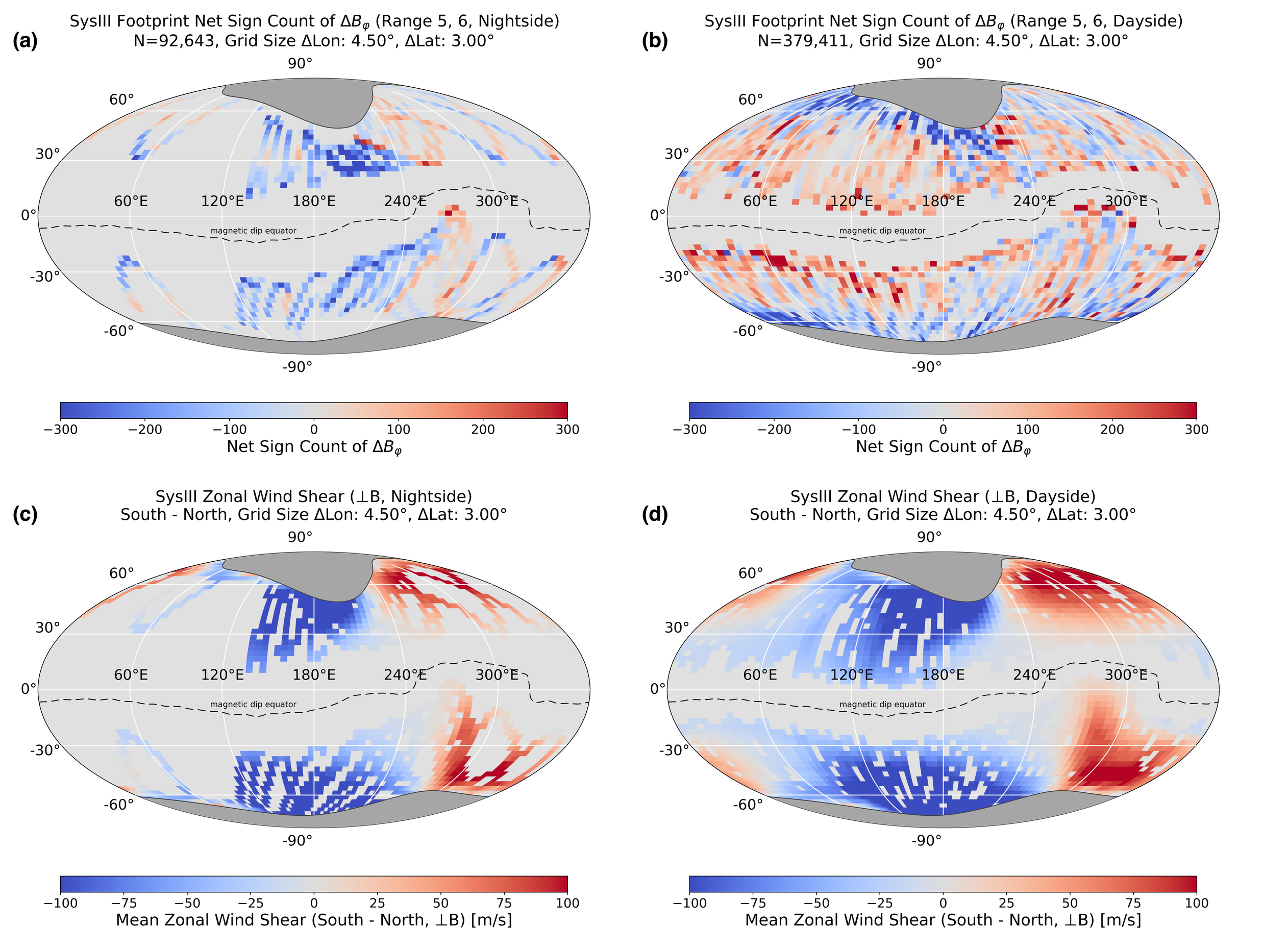}
\caption{Observed System III ionospheric footprint net sign count of $\Delta B_\varphi$ for Juno MAG ranges 5--6: (a) nightside and (b) dayside. JTIM southern-minus-northern eastward zonal wind shear: (c) nightside and (d) dayside. The observed polarity broadly agrees with the modeled wind shear on the nightside but differs on the dayside.}
\label{fig:JTIM_compare}
\end{figure*}

The inferred current system suggests a hemispheric asymmetry in zonal winds shear that varies with local time. To further investigate this interpretation, we compare the observed $\Delta B_\varphi$ polarity distributions with the hemispheric zonal wind shear predicted by JTIM. Figure~\ref{fig:JTIM_compare} compares the observed $\Delta B_\varphi$ polarity distributions with the hemispheric zonal wind shear predicted by JTIM at 400-500 km altitude for both the nightside and dayside sectors. The modeled thermospheric winds are stable with local time, consistent with observations of ionospheric vertical structure at Jupiter, which show minimal local time variations in electron density \cite{agiwal_unraveling_2025,moore_photochemistry_2026}. However, Figure \ref{fig:Bphi_mollweide} and Figure \ref{fig:Bphi_zrho} reveals strong local time variations in the observed $\Delta B_\varphi$ polarity, suggesting that wind-shear driven FACs alone cannot explain the observed sub-Io magnetic field perturbations. Figure \ref{fig:JTIM_compare} panels (a) and (b) show the nightside and dayside net sign count distributions, respectively, of the $\Delta B_\varphi$ mapped to ionospheric footprints for Juno MAG Nominal Dynamic Range~5--6 \cite{connerney_juno_2017}. Range~1--4 corresponds to magnetic field strengths below approximately $1.0\times10^5$~nT, whereas Range~5--6 corresponds to stronger-field regions above approximately $1.0\times10^5$~nT, generally sampling locations closer to Jupiter and at lower mapped latitudes. Panels (c) and (d) show the corresponding nightside and dayside JTIM-predicted east-west neutral wind difference between the southern and northern hemispheres along the magnetic field lines.

On the nightside (Figures~\ref{fig:JTIM_compare}a and \ref{fig:JTIM_compare}c), the observed $\Delta B_\varphi$ polarity distribution is broadly consistent with the JTIM-predicted wind difference between the southern and northern hemisphere ionosphere footprint. This agreement supports the interpretation that the nightside IHFAC system is mainly generated by Jovian thermospheric wind-driven dynamo processes. Unlike the nightside structure, the dayside (Figures~\ref{fig:JTIM_compare}b and \ref{fig:JTIM_compare}d) low-latitude $\Delta B_\varphi$ distribution appears significantly less longitudinally organized and shows large deviations from the modeled wind shear pattern. This discrepancy suggests that additional physical processes, not fully captured by the current JTIM framework, contribute to the observed residual magnetic field on the dayside. 

Current JTIM simulations adopt a prescribed, time-independent electric field \cite{cowley_simple_2005}, without incorporating self-consistent electrodynamic feedback or wind-driven dynamo electric fields at middle and low latitudes. While this approach provides a useful baseline for evaluating thermospheric wind-driven processes, it may not fully capture the coupling between thermospheric winds, ionospheric electric fields, and plasma transport at mid and low latitude. At Earth, variations in thermospheric winds and geomagnetic activity can modify the wind-driven ionospheric current system and associated IHFACs \cite{richmond_ionospheric_1979,yamashita_seasonal_2002}. At Saturn, ring-derived material can alter ionospheric composition and electron densities, potentially affecting ionospheric conductivity and current closure \cite{odonoghue_domination_2013,waite_chemical_2018,agiwal_current_2024}. These examples suggest that, at Jupiter, additional ionospheric processes, ring--planet coupling, and temporal variations in auroral energy input could modify the distribution and strength of mid- to low-latitude IHFACs. In particular, solar-modulated processes such as the equatorial plasma fountain may introduce local-time and longitudinal variations in ionospheric plasma distributions \cite{agiwal_unraveling_2025}, although their role in Jupiter's ionosphere remains uncertain. Such processes, together with the electrodynamic feedback not yet represented in JTIM, could contribute to the discrepancy between the observed dayside magnetic perturbations and the modeled wind shear. Further observations and self-consistent electrodynamic modeling are needed to assess their relative contributions and understand the pronounced day--night asymmetry of Jupiter's mid- to low-latitude IHFAC system.

\section{Summary}
To summarize, we analyze Juno MAG observations from the first 62 perijove passes and identify a distinct spatial distribution of the azimuthal residual magnetic field $\Delta B_\varphi$ within 5~R$_\mathrm{J}$, characterized by a latitudinal "sandwich" structure and a pronounced day--night asymmetry. Using new models of Jupiter's thermospheric winds, we show that the observed $\Delta B_\varphi$ polarity pre-7 h local time is consistent with an interhemispheric field-aligned current system arising from a wind-driven dynamo. On the dayside, however, the FACs reorganize dramatically, shifting from longitude-controlled to latitude-controlled structure and deviating substantially from wind-driven predictions, indicating that processes beyond those captured in the current version of JTIM govern dayside currents.
These results provide new insights into ionosphere-thermosphere coupling at Jupiter and represent the first evidence of field-perpendicular ion-neutral coupling in Jupiter's ionosphere, complementing recent observations of field-parallel ion-neutral coupling \cite{agiwal_unraveling_2025,moore_photochemistry_2026}. They raise a critical question: what drives the dayside IHFAC at Jupiter? The coincidence of the latitude-organized dayside structure with Jupiter's main ring footprint suggests that ring-ionosphere coupling may warrant further study at Jupiter. These results highlight the importance of local-time-dependent and longitude-dependent electrodynamics Jupiter's upper atmosphere and M-I coupling in the innermost magnetosphere. 

\appendix

\section{Orbit Lists Used in this Study}

The GBS orbits excluded from analysis are PJ: 1, 12, 14, 19, 24, 26, 36, 37, 38, 39, 41, 42, 52, 57, and 60. 

The corresponding non-GBS orbits used in the IHFAC fitting are PJ: 3, 4, 5, 6, 7, 8, 9, 10, 11, 13, 15, 16, 17, 18, 20, 21, 22, 23, 25, 27, 28, 29, 30, 31, 32, 33, 34, 35, 40, 43, 44, 45, 46, 47, 48, 49, 50, 51, 53, 54, 55, 56, 58, 59, 61, and 62.

For a subset of the non-GBS orbits, the observed $\Delta B_\varphi$ cannot be adequately described by our simple IHFAC model. These orbits are PJ: 3, 4, 27, 31, 32, 43, 44, 45, 46, 49, 51, and 59. We therefore did not include these orbits in the forward-modeled IHFAC field interpretation and figures.
%



%
%

\section*{Open Research Section}
The Juno MAG data used in this study are publicly available through the NASA Planetary Data System \cite{jep_connerney_juno_2024}.

\section*{Conflict of Interest declaration}
The authors declare there are no conflicts of interest for this manuscript.

\begin{acknowledgments}
This research was supported by the NASA Juno mission.
\end{acknowledgments}
%
%

\bibliography{references}

%
%
%
%
%

\end{document}